\PassOptionsToPackage{table}{xcolor}
\documentclass[acmsmall,screen,nonacm]{acmart}
\AtBeginDocument{
  }

\usepackage{amssymb}
\usepackage{array}
\usepackage{tikz}
\usepackage{listings}
\usepackage{lstnix}
\usepackage[export]{adjustbox}
\usepackage{siunitx}
\usepackage{xspace}
\usepackage{pdfcomment}
\usepackage{cleveref}

\makeatletter
\AddToHook{enddocument/afterlastpage}{
  \begingroup
    \ulp@cnt\z@
    \immediate\openout\ulp@out=\jobname.upb\relax
    \begingroup
      \makeatletter
      \InputIfFileExists{\jobname.upa}{}{}
    \endgroup
    \immediate\closeout\ulp@out
  \endgroup
}
\makeatother

\def \archiveSize {56\xspace}
\def \LLMEvalSize {110\xspace}
\def \LLMEvalPercent {54.5\%\xspace}

\newcolumntype{P}[1]{>{\centering\arraybackslash}p{#1}}
\newcolumntype{L}[1]{>{\raggedright\arraybackslash}p{#1}}
\newcolumntype{R}[1]{>{\raggedleft\arraybackslash}p{#1}}

\begin{document}

\title[NICE: Declarative and Machine-Checkable Vulnerability Reproduction]{NICE: A Framework for Declarative and Machine-Checkable
  Vulnerability Reproduction}

\author{Minh-Lu\^an Nguyen}
\email{minh-luan.nguyen@telecom-sudparis.eu}
\affiliation{
  \institution{SAMOVAR, T\'el\'ecom SudParis, Polytechnic Institute of Paris}
  \city{Palaiseau}
  \country{France}}

\author{Olivier Levillain}
\email{olivier.levillain@telecom-sudparis.eu}
\affiliation{
  \institution{SAMOVAR, T\'el\'ecom SudParis, Polytechnic Institute of Paris}
  \city{Palaiseau}
  \country{France}}

\author{Julien Malka}
\email{julien.malka@telecom-paris.fr}
\affiliation{
  \institution{LTCI, T\'el\'ecom Paris, Polytechnic Institute of Paris}
  \city{Palaiseau}
  \country{France}}

\author{Stefano Zacchiroli}
\email{stefano.zacchiroli@telecom-paris.fr}
\affiliation{
  \institution{LTCI, T\'el\'ecom Paris, Polytechnic Institute of Paris}
  \city{Palaiseau}
  \country{France}}

\author{Th\'eo Zimmermann}
\email{theo.zimmermann@telecom-paris.fr}
\affiliation{
  \institution{LTCI, T\'el\'ecom Paris, Polytechnic Institute of Paris}
  \city{Palaiseau}
  \country{France}}

\renewcommand{\shortauthors}{Nguyen et al.}

\begin{abstract}
  Reproducing software vulnerabilities is fundamental to security researchers, open-source maintainers, and educators. Yet, vulnerabilities are often hard to reproduce, and even when they can be reproduced once, recreating a software environment where the vulnerability can be exploited becomes harder and harder over time. We present \emph{NICE}, the NIx CvE reproduction framework, which uses declarative recipes to build and automatically validate vulnerable environments. In NICE, a reproduced CVE comprises one or more NixOS virtual machine configurations, a scripted exploitation scenario, and machine-checkable assertions that provide factual evidence of exploitation. This design facilitates sharing, validation, review, and long-term reproducibility.

  Building on NICE, we construct a public archive of \archiveSize\ reproduced vulnerabilities spanning multiple CWE categories, attack vectors, and target types (user-space, system software, kernel, and graphical applications), combining manually authored recipes with recipes produced by an LLM-based automated pipeline and merged after human review. We evaluate the feasibility of the automated pipeline on \LLMEvalSize\ real-world CVEs, achieving fully automated reproduction in \LLMEvalPercent\ of the cases, at an average cost of \$0.73 per CVE. The evaluation demonstrates that NICE provides a practical target for LLM-assisted recipe generation, while also identifying where human refinement or additional automation remains necessary.

  NICE is applicable to security education and training (e.g., creating cyber ranges), but also to vulnerability reporting, where its reproducibility and reviewability properties can make reports easier to audit and verify.\end{abstract}



\maketitle

\section{Introduction}

Reproducing software vulnerabilities is a fundamental activity in vulnerability research and practice. It enables researchers and practitioners to investigate root causes, assess exploitability, and validate patches under controlled conditions. Reproducible environments also support security education and training by allowing vulnerabilities to be studied safely and repeatedly without affecting production systems.

However, in today's practice, vulnerability reproduction is often fragile, time-consuming, or altogether infeasible. Vulnerabilities that cannot be reliably reproduced tend to receive less scrutiny and may remain unresolved or poorly understood~\cite{HackersVsTesters}.
Empirical evidence highlights the severity of this problem: a large-scale study of 368 memory-error vulnerabilities reported that, despite over 3,600 hours of manual effort, only 54.9\% of vulnerabilities could be successfully reproduced~\cite{CrowdSec}.
The study identifies missing, ambiguous, or incomplete information in vulnerability reports as a major barrier, particularly regarding build configurations, dependencies, and execution environments. These findings suggest that reproducibility failures are often not inherent to vulnerabilities themselves, but rather stem from inadequate mechanisms for capturing and reconstructing the software environments in which vulnerabilities manifest.

Functional package management (FPM) provides a promising foundation for addressing these challenges. Its implementations, such as Nix~\cite{Dolstra2006} and Guix~\cite{guix}, offer a declarative model to specify dependencies, which has been shown to enable reproducibility of software environments over time~\cite{NixSpaceTime}.
FPM-based approaches have already been used to improve reproducibility in domains such as scientific computing, bioinformatics, chemistry, and IoT systems~\cite{kowalewski2022sustainable, SelfScalingClusters2020, 10.1093/gigascience/giaa121, AutomatedIoTAppSanti}.
Meanwhile, standards in vulnerability reporting and reproduction remain ad-hoc.

In this paper, we bridge the gap between reproducibility and vulnerability research by introducing \emph{NICE},~\footnote{To be pronounced as the city of Nice.}
the NIx CvE reproduction framework, which leverages FPM to create reproducible vulnerable software environments and automate the validation of their exploitability.
NICE allows the expression of vulnerability reports into declarative, executable artifacts that precisely specify (1) the software environment in which a vulnerability occurs and (2) the procedures required to demonstrate exploitability. By making these artifacts explicit and machine-testable, NICE reduces manual effort, minimizes configuration errors, and enables independent verification of exploitability claims.

Our framework enables packaging hands-on security labs and cyber range exercises as lightweight, reviewable environments that can be shared and redeployed easily. Accompanying automated validation tests help ensure that these labs continue to behave as intended, which is useful for instructors and researchers alike. NICE can also strengthen vulnerability reporting and triage by providing machine-checkable evidence of exploitation for reviewers and maintainers, increasing confidence and reducing ambiguity during coordinated disclosure and patch validation.

This paper makes the following contributions:
\begin{itemize}
  \item \textbf{NICE}, a framework comprising a methodology, documentation, a CLI tool, and a library, that allows expressing vulnerability reproduction as declarative, machine-checkable Nix recipes combining NixOS virtual machine configurations, exploitation scenarios, and reusable assertion blocks (\Cref{sec:framework}).
  \item \textbf{A public archive of reproduced vulnerabilities}, combining manually authored recipes with recipes produced by an LLM-based automated pipeline and validated through human review, currently comprising \archiveSize\ recipes spanning diverse CWE categories, attack vectors, and target types (\Cref{sec:archive}).
  \item \textbf{An empirical study of LLM-based automated recipe generation}, evaluating the feasibility, success rate, and failure modes of using autonomous agents to grow the NICE archive, as well as the reliability of their own success and failure verdicts. Our experiments show that NICE recipe generation can be fully automated in \LLMEvalPercent\ of the cases and with an average cost of \$0.73 per CVE. (\Cref{sec:llm-evaluation}).
\end{itemize}

This paper is structured as follows. \Cref{sec:background} provides necessary background on NICE's technical foundations: NixOS and its test framework. \Cref{sec:framework} presents the NICE framework, its design objectives, its technical design, and its usage workflow. \Cref{sec:archive} presents the NICE archive of reproduced vulnerabilities. \Cref{sec:llm-evaluation} presents our evaluation of the LLM-based automated recipe generation. \Cref{sec:discussion} discusses the implications of this work. Before concluding, \Cref{sec:related} reviews related work in vulnerability reproduction and FPM-based reproducibility.

\section{Background}
\label{sec:background}

In this section we provide an overview of the main technical building blocks we used to build the NICE framework, namely: Nix, NixOS, and the NixOS test framework.

\paragraph{Nix~\cite{Dolstra2006}} A functional package manager that declaratively specifies software environments and supports their reproducible reconstruction~\cite{NixSpaceTime}. Its community-maintained package collection, Nixpkgs, contains over \num{80,000} packages,\footnote{According to Repology {https://repology.org/} as of January 2026.} providing broad coverage for vulnerable software and dependencies. Nix also supports pinning environments to specific Nixpkgs revisions; its Git history extends to 2003, enabling reconstruction of historical software environments.

\paragraph{NixOS~\cite{nixos}} A Linux distribution built on top of the Nix package collection. A key characteristic of NixOS is its module system, which allows the entire system configuration (including installed packages, system services, user accounts, network settings, etc.) to be specified declaratively in a single configuration file written in the Nix programming language. This configuration can then be evaluated by Nix to apply changes to a running system or to generate system images (e.g., virtual machines), while ensuring that the same configuration yields the same software environment across machines and over time.

\paragraph{The NixOS test framework~\cite{NixOSTests}} An automated integration testing framework for the NixOS distribution. It allows developers to specify one or more NixOS virtual machine configurations together with a test scenario that orchestrates their execution and interactions. The test framework provides a programmable Python interface to instrument the execution of virtual machines to perform specific actions and assert properties of the running systems such as whether services are running or specific commands produce expected outputs.

\section{NICE Framework}
\label{sec:framework}

\begin{figure*}
  \centering
  \includegraphics[width=0.9\linewidth]{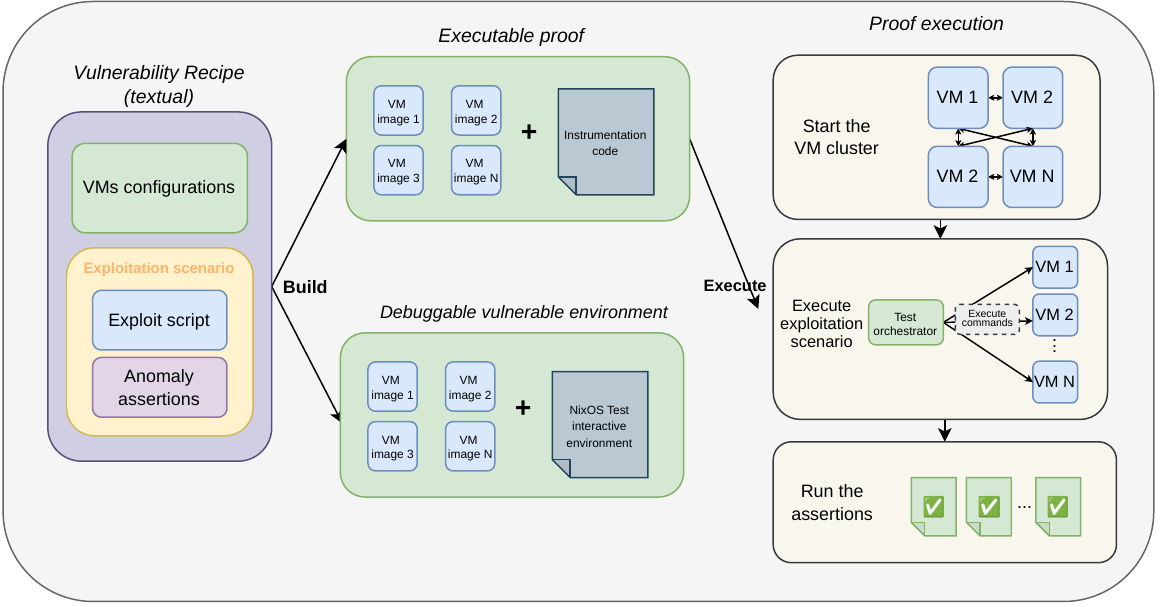}
  \caption{Overview of the NICE framework for vulnerability reproduction using NixOS.}
  \Description{Block diagram of the NICE framework: a vulnerability recipe is built either into a debuggable virtual machine environment or into a self-contained executable proof of exploitation.}
    \label{fig:framework-overview}
\end{figure*}

In this section, we present the NICE framework for reproducing and automatically validating software vulnerabilities.
We first describe the motivation that led to the design of the framework, then we describe components that enable reproducing vulnerabilities of interest. Finally, we present a workflow that can be followed to combine those components to do so.

\subsection{Design objectives}
\label{sec:design}

We begin by outlining the design objectives that guided the development of the NICE framework. The design space of vulnerability reproduction tools is broad and heterogeneous, encompassing a wide range of approaches and assumptions, discussed in \Cref{sec:related}. Existing tools vary significantly in scope and focus: some emphasize automated vulnerability packaging—where a vulnerable environment is generated from a CVE identifier, with minimal human intervention—while others target specific classes of vulnerabilities (e.g., memory corruption, web application flaws) or specific execution environments.

Our work focuses on reducing the friction associated with the \emph{validation} phase of vulnerability reproduction. Rather than aiming to fully automate environment generation or exploit development, we prioritize making it easy for users to verify that a vulnerability is present and exploitable under well-defined conditions. In particular, our framework is designed around the following objectives:
\begin{itemize}

\item \textbf{Lightweight sharing.} Sharing a vulnerability reproduction should not require distributing large or opaque binary artifacts such as pre-built containers, virtual machine images, or disk snapshots. Instead, vulnerability descriptions should be lightweight and portable. When a vulnerability is not inherently tied to a specific processor architecture, this approach also avoids artificially constraining reproductions to a single architecture.

\item \textbf{Push-button validation.} Once a vulnerability is packaged, users should be able to trigger the validation of its exploitability with minimal effort, executing a single command, without requiring manual configuration, environment setup, or in-depth knowledge of the underlying exploit.

\item \textbf{Ease of review.} The framework should facilitate human understanding and independent verification. Users seeking to convince themselves that an unintended or improper behavior occurs should be able to do so without having to study a complex exploit in detail. Instead, the evidence of vulnerability presence should be explicit, observable, and easy to reason about.

\item \textbf{Reproducibility over time.} None of the properties above should degrade as software ecosystems evolve. A packaged vulnerability should remain verifiable with the same level of effort and confidence years after its initial creation, despite the continuous release of new software libraries, tools, package managers, and execution environments.
\end{itemize}

\noindent
In addition to the above usability-oriented goals, we also design the framework to be \emph{versatile} across a broad range of vulnerabilities. Specifically, NICE aims to support:
\begin{itemize}

\item \textbf{Diversity of vulnerability classes.} Vulnerabilities spanning a wide range of CWE categories, rather than focusing on a single class or exploit primitive.

\item \textbf{Diversity of attack vectors.} Vulnerabilities whose exploits rely on a large variety of attack vectors, such as network-based exploits (e.g., remote code execution), local exploits (e.g., privilege escalation), resource exhaustion (e.g., denial of service), etc.

\item \textbf{Diversity of targets.} Vulnerabilities affecting different categories of software, from kernel to userland, and from command-line utilities to client/server systems and graphical applications.

\item \textbf{Temporal coverage.} Vulnerabilities drawn from a large temporal span, including older vulnerabilities. This is important both for teaching and research applications as older vulnerabilities are often easier to instrument for students, and new vulnerability detection tools should not be tested exclusively on recent vulnerabilities.
\end{itemize}

\subsection{Technical description}

\Cref{fig:framework-overview} summarizes NICE's technical components, showing how a vulnerability recipe specifies a reproducible, runnable vulnerable software environment and an executable proof that automatically validates its exploitability.

\subsubsection{Vulnerability recipe}

The NICE framework builds upon the NixOS test framework to automate the execution and validation of software vulnerabilities.
NICE allows authors to express \emph{vulnerability recipes} as Nix expressions that declaratively specify all artifacts required to reproduce and validate a vulnerability.

A vulnerability recipe consists of two main components: the \emph{virtual machine configurations} and the \emph{exploitation scenario}.
Together, these elements encode how to deploy the vulnerable environment, how to exploit the vulnerability in it, and how to determine whether exploitation was successful.

\paragraph*{Virtual machine configurations}
The first part of a vulnerability recipe specifies one or more \emph{virtual machine configurations}. Each configuration fully describes a NixOS system, including but not limited to: the Linux kernel version, installed packages, running services, system users, and network settings required for the vulnerability to manifest itself.

Recipes may describe a single VM or a cluster of interacting VMs, depending on the vulnerability (e.g., client/server or multi-host scenarios).
Similarly, the attack vector may require multiple users with different privileges to exist (e.g., for a local privilege escalation scenario).
The VM configurations use the NixOS module system introduced in \Cref{sec:background}, allowing an external reader to understand the software environment deployed in the virtual machines by reading the recipe.

\paragraph*{Exploitation scenario}

The second part of the vulnerability recipe defines an \emph{exploitation scenario}, which specifies how the vulnerability is exercised and how its successful exploitation is detected. Exploitation scenarios are written in Python and use the NixOS test framework API, which provides primitives for orchestrating virtual machines, such as booting systems, waiting for \texttt{systemd} services to become available, checking network connectivity, and executing arbitrary commands within virtual hosts.

Using this interface, the exploitation scenario encodes an \emph{exploit script} that performs the actions required to trigger the vulnerability. The scenario then validates the outcome of the exploit by checking whether the system has transitioned into an unauthorized state. Such checks may include checking that a process has obtained elevated privileges, or observing the creation or modification of files in privileged locations.

NICE uses publicly available proof-of-concept exploits to exercise vulnerabilities. Simple PoCs, such as shell scripts or commands using existing Linux tools, can be invoked directly from the exploitation scenario. Exploits requiring specific toolchains or libraries are packaged as Nix derivations so that their dependencies and build instructions are captured alongside the vulnerable environment.

To improve modularity and reuse, the NICE library encapsulates common validation patterns into \emph{assertion blocks}. An assertion block is a piece of code that implements a specific validation check and can be used in exploitation scenarios as needed. This abstraction improves readability by separating high-level validation logic from low-level exploit mechanics: \textbf{external readers do not need to reason about the details of exploit execution, but only acknowledge that a given unauthorized system state has been reached}, thus helping achieve the ``Ease of review'' objective. Assertion blocks also enhance extensibility, as new blocks can be added to support additional vulnerability classes. The assertion blocks currently provided by NICE are summarized in \Cref{table:assertion-blocks}.

\begin{table*}
  \caption{Assertion blocks included in the NICE library. Each block checks, on a given machine, that the stated condition holds.}
  \centering
  \small
  \begin{tabular}{@{}lp{8.5cm}@{}}
  \toprule
  \textbf{Assertion block} & \textbf{Checked condition} \\
  \midrule
  \texttt{assert\_service\_log\_contains} & a systemd service's journal contains a given message \\
  \texttt{assert\_root\_privileges} & a user has root privileges \\
  \texttt{assert\_file\_exists} & a file exists \\
  \texttt{assert\_file\_contains} & a file contains a given content \\
  \texttt{assert\_core\_dump\_exists} & a core dump exists for a given process and signal \\
  \texttt{assert\_exact\_execution\_time} & a command runs for the expected duration \\
  \texttt{assert\_file\_size\_equals} & a file has the expected size in bytes \\
  \texttt{assert\_cpu\_usage\_high} & CPU usage exceeds a threshold (DoS, infinite loop) \\
  \texttt{assert\_memory\_usage\_high} & Memory usage exceeds a threshold (DoS, infinite loop) \\
  \texttt{assert\_screen\_text} & The screen contains a given text \\
  \bottomrule
  \end{tabular}
  \label{table:assertion-blocks}
\end{table*}

\subsubsection{Building and running vulnerable environments and executable proofs}

The NICE CLI supports two complementary ways of building and running a vulnerability recipe.

First, the CLI can build and run a debuggable environment that a user can log into to interact with the VM(s) in real time. This can be useful for debugging purposes when constructing the vulnerability recipe for the first time, but also, later on, for research or teaching purposes (e.g., a student having to learn how to manually exploit the vulnerability).

Second, the CLI can build and run a fully self-contained executable proof of the vulnerability's existence. This includes both the set of NixOS virtual machine images corresponding to the specified VM configurations and the instrumentation code derived from the exploitation scenario. The instrumentation code includes all the scaffolding code used by the test framework to communicate and execute commands on the virtual machine through the \texttt{qemu} monitoring features as well as all the recipe-specific code to run the test scenario, that is the exploit logic and the assertions that will later be executed during automated testing. With a single CLI command, NICE can build the environment, boot the VM cluster, execute the exploitation scenario, and evaluate the assertions.

During execution, the host machine acts as an orchestrator, coordinating interactions between VMs and driving the exploit by reading system state and issuing commands through the \texttt{qemu} socket. If all assertions hold, the vulnerability is considered successfully reproduced; otherwise, execution fails, indicating that the exploit did not trigger the expected unauthorized system state under the specified conditions. NICE reports explicit success or failure status together with execution traces that can be inspected for debugging and independent verification. Proofs can therefore be executed fully unsupervised, including in continuous integration environments.

This transformation from the vulnerability recipe into the runnable environment (be it the debuggable environment or the one from the automated executable proof) is \textbf{reproducible in space and in time}: it is deployable on different machines (with the Nix toolchain as the only external dependency) and is future-proof, as empirically validated in the past for general Nix packages~\cite{malka:hal-04913007}.
Given this step always yields the same environment, \textbf{sharing the lightweight proof recipe is sufficient to reproduce the vulnerability} in the future and elsewhere---one does not need to share the large binary VM images produced by the build. Therefore, we achieve both our ``Lightweight sharing'' and ``Reproducibility over time'' objectives.

\begin{figure*}
    \centering
    \includegraphics[width=1.0\linewidth]{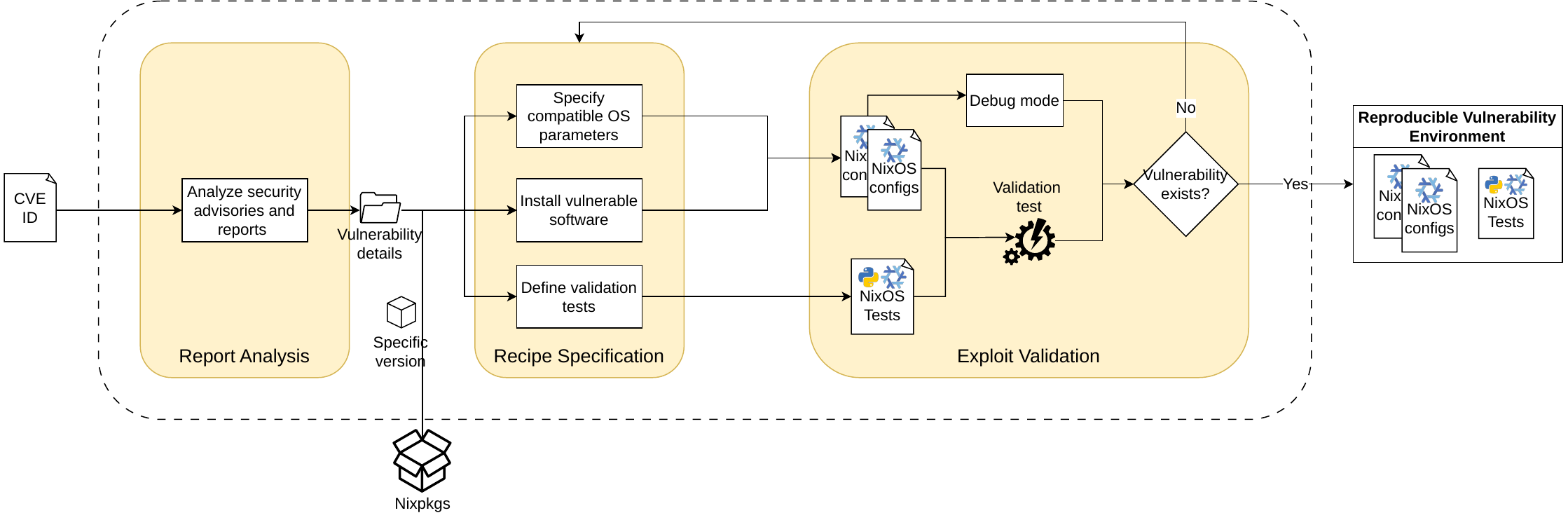}
    \caption{Workflow for vulnerability reproduction using our NICE framework.}
    \Description{Three-stage workflow: report analysis, recipe specification, and exploit validation.}
    \label{fig:workflow-overview}
\end{figure*}

\begin{figure}[!t]
  \begin{adjustbox}{max width=0.5\linewidth}
    \begin{lstlisting}[language=Nix, escapechar=|]
{ config, pkgs, ... }:
let
  opensslPkg = (import (builtins.fetchTarball {
    url = "https://github.com/.../139ff...tar.gz";
    sha256 = "sha256:1s8zwlzvany97xmv7qcc...";
  }) { system = "x86_64-linux"; }).openssl;  |\boxed{1}|
in
{
  environment.systemPackages = with pkgs;
    [ opensslPkg ];
  networking.firewall.allowedTCPPorts = [ 443 ]; |\boxed{2}|
  systemd.services.opensslSetup = { |\boxed{3}|
    wantedBy = [ "multi-user.target" ];
    path = [ opensslPkg ];
    script = ''
      openssl req -x509 -newkey rsa:2048 \
        -out cert.pem -keyout key.pem
      openssl s_server -key key.pem \
        -cert cert.pem -accept 443 -www
    '';
  };
}
    \end{lstlisting}
  \end{adjustbox}
  \caption{OpenSSL server VM configuration for the Heartbleed vulnerability (simplified).}
\Description{Listing of the NixOS configuration of the Heartbleed server: OpenSSL 1.0.1e is fetched from a pinned Nixpkgs revision, port 443 is opened, a systemd service starts an OpenSSL TLS server, and a root account is created.}
  \label{fig:nixos-configuration-heartbleed}
\end{figure}

\subsection{Vulnerability reproduction workflow}
\label{sec:workflow}

In this section, we describe the typical workflow a reproducer would follow to turn a vulnerability report and a public proof-of-concept exploit into a NICE recipe, from analyzing the report to configuring the environment and validating exploitation.
NICE's documentation aims to be usable by both humans and agents, thus enabling LLM-automated reproduction, as evaluated in \Cref{sec:llm-evaluation}.
Figure~\ref{fig:workflow-overview} shows the NICE workflow divided into three stages: \emph{Report Analysis}, \emph{Recipe Specification}, and \emph{Exploit Validation}.

In the following, we use Heartbleed (CVE-2014-0160) as a running example of each stage.

\paragraph{Report analysis.}
We start from public vulnerability reports and advisories (e.g., NVD/MITRE entries and technical write-ups) and extract three main pieces of information:
\begin{itemize}
  \item \textbf{Affected software and version range.} For Heartbleed, OpenSSL versions \texttt{1.0.1} through \texttt{1.0.1f} are affected, and version \texttt{1.0.1g} contains the fix.
  \item \textbf{Attack vectors and roles.} Heartbleed is remotely triggerable against a TLS entity. Therefore, a possible exploitation scenario involves a \emph{server} running the vulnerable library, a \emph{client} that sends a private message to the server, and an \emph{attacker} that sends a crafted heartbeat request to leak the client's message from the server's memory.
  \item \textbf{Operational preconditions.} We record requirements such as system configurations, relevant services, user privileges, network exposure, and whether proofs of concept (PoCs) assume specific conditions. In Heartbleed, the server must expose its TLS endpoint for remote connections.
\end{itemize}

Once we know which software versions we need in the environment, we need to identify them among historical revisions of the Nixpkgs repository.
We use tools such as \emph{nixpkgs-multiverse}\footnote{Available online at: \url{https://nixmultiverse.com/}} to find which Nixpkgs revisions contain the relevant OpenSSL versions. Using this tool, we learn that revision \texttt{139ff} of Nixpkgs contains version \texttt{1.0.1e}, which is affected.

\paragraph{Recipe specification.}

We then express the vulnerable environment as a vulnerability recipe. Two decisions need to be made:

\smallskip
\noindent\textbf{(i) Decide the number and roles of machines.}
For Heartbleed, we choose to rely on 3 VMs, one for each role in a typical attack scenario: a server, a client, and an attacker.

\smallskip
\noindent\textbf{(ii) Select a pinning strategy for Nixpkgs.}
Depending on how tightly coupled a vulnerability is to its ecosystem, one of the two following strategies might be preferable:
\begin{itemize}
  \item \emph{Snapshot pinning}: pin the full NixOS system to a historical Nixpkgs revision when integration details matter (e.g., kernel/userland alignment, older infrastructures, brittle dependency stacks). This ensures that the vulnerable component is built and installed in the same software environment as at the time of vulnerability discovery.
  \item \emph{Component pinning}: fetch only the vulnerable component from an older Nixpkgs revision while pinning the rest of the NixOS system to a recent version. Using a modern NixOS version means that we will get all the NixOS test framework features we need for the setup, provided that the vulnerable component remains compatible with its surrounding environment. In addition, this strategy reduces the risk of unintentionally reintroducing unrelated, known vulnerabilities elsewhere in the stack.
\end{itemize}

Figure~\ref{fig:nixos-configuration-heartbleed} shows a simplified version of the NixOS configuration for the OpenSSL server in Heartbleed.
Given that OpenSSL is a self-contained library with limited system dependencies, we use component pinning.
We fetch OpenSSL \texttt{1.0.1e} from an older Nixpkgs revision \boxed{1} while building the rest of the system with a recent Nixpkgs version.

The NICE library provides a minimal configuration template for every VM, consisting of a base NixOS configuration with a few essential packages (e.g., \texttt{bash}, \texttt{coreutils}) and a default root account. From here, the recipe author can add the specific packages and services required for the vulnerability environment.

A Heartbleed vulnerable environment requires an OpenSSL TLS server installed \boxed{3} and a port opened \boxed{2} to allow remote connections. The TLS server is started on boot using a \texttt{systemd} service.
The client and the attacker VMs are simpler, as they only require packages such as \texttt{curl} for the client's connection and the exploit setup for the attacker.

\begin{figure*}[!t]
    \centering

    \begin{minipage}[b]{0.45\textwidth}
        \vspace{0pt}
        \centering

        \begin{adjustbox}{
            max width=\linewidth,
            max totalheight=\textheight
        }
        \begin{lstlisting}[language=Nix]
import assertion_blocks as ab

dump_file = "/tmp/dump.bin"
dump_length = 16384
secret_key = "752148647812547924"

server.start()
client.start()
server.wait_for_unit("multi-user.target")
server.wait_for_unit("opensslSetup.service")
server.wait_for_open_port(443) <@\boxed{4}@>
client.wait_for_unit("multi-user.target")

message_data = f"Hey! ...
The code for that room is {secret_key}."
client.execute(f"curl -b message='{message_data}'
 https://server:443") <@\boxed{5}@>

attacker.start()
attacker.wait_for_unit("multi-user.target")
attacker.execute(f"attack-script -u server -p 443
-f {dump_file} -l {dump_length}") <@\boxed{6}@>

ab.check_file_contains(
    attacker, dump_file, secret_key) <@\boxed{7}@>
        \end{lstlisting}
        \end{adjustbox}

        \captionof{figure}{
            NixOS test for the Heartbleed vulnerability (simplified).
        }
        \Description{
            Listing of the NixOS test for Heartbleed: it declares
            the server, client and attacker nodes and a Python test
            script running the exploit and assertion blocks.
        }
        \label{fig:nixos-test-heartbleed}
    \end{minipage}
    \hfill
    \begin{minipage}[b]{0.50\textwidth}
        \vspace{0pt}
        \centering

        \includegraphics[
            width=0.8\linewidth
        ]{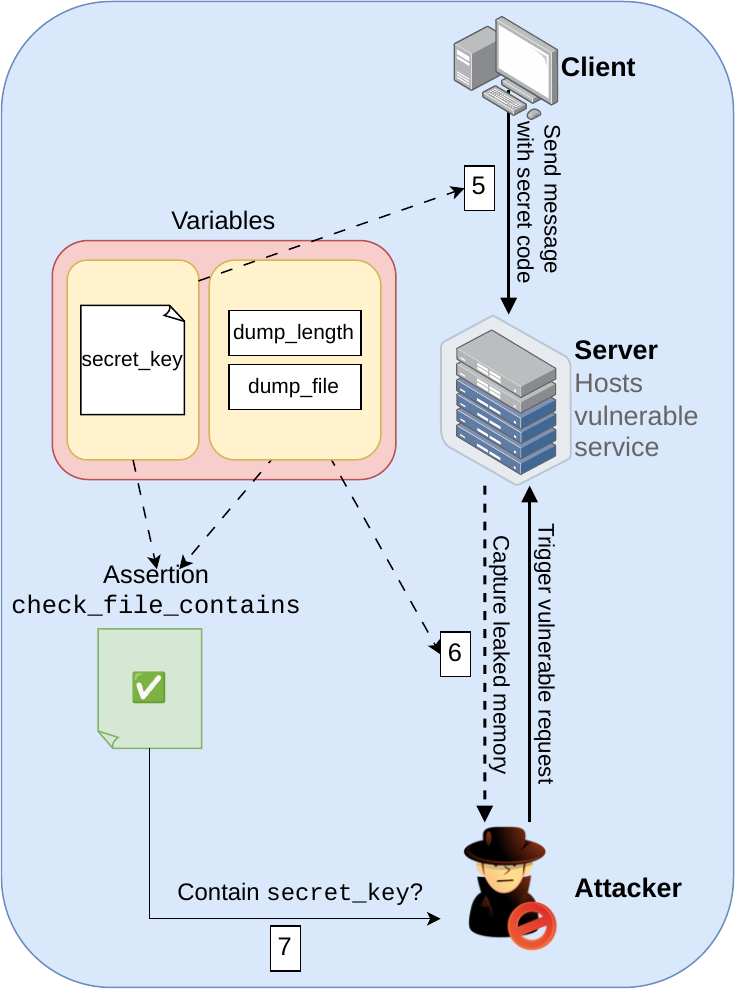}

        \captionof{figure}{
            Test orchestration diagram for the Heartbleed vulnerability.
        }
        \label{fig:heartbleed-test-diagram}
    \end{minipage}

\end{figure*}

\paragraph{Exploit validation.}
In the final stage, we encode the exploitation procedure and its validation checks in the recipe's test script. For Heartbleed, the test starts the three VMs, waits for the TLS service to become available, sends a private message from the client, and runs the attacker's PoC. It then checks whether the leaked data contains the client's secret. Figure~\ref{fig:nixos-test-heartbleed} shows the corresponding code, which we explain below. The debuggable environment can be used to troubleshoot these steps interactively.

Here we set up a validation test for the scenario described earlier to automatically confirm the presence of the Heartbleed vulnerability in the environment. The test script defines the procedure with the help of built-in functions provided by the NixOS test framework and helper functions from the NICE framework. In this case, we first initialize the VM cluster and ensure the vulnerable service is running and reachable \boxed{4}. Then, the client VM sends a private message containing a secret key to the server \boxed{5}. Next, we execute the exploit script on the attacker node \boxed{6} (installed in the attacker's system packages), which sends a malformed heartbeat request that causes the server to leak a fixed, attacker-controlled amount of memory into a file. Although the leaked memory contents are unpredictable, the requested payload length is constant, making the leak size deterministic. Finally, \boxed{7} validates the exploit's success by checking the response and the presence of secret key material in the leaked data. Figure~\ref{fig:heartbleed-test-diagram} illustrates how the test orchestrates the interactions between the three VMs.

A key good practice is to keep test parameters coherent: input values used to configure the PoC (e.g., output path and expected length) \boxed{5} are shared with the corresponding assertion block \boxed{7} and the exploit invocation \boxed{6}. This strengthens the correlation between the exploit action and the observed effect, making the validation more convincing.

\section{The NICE Archive}
\label{sec:archive}

The NICE archive currently comprises \archiveSize\ CVEs, selected as a purposeful sample covering heterogeneous reproduction settings and the versatility properties defined in \Cref{sec:design}. For each CVE, we produced at least one \emph{VM configuration} and an \emph{exploitation scenario}.

We built the archive from recipes obtained through a variety of methods.
Some of them were entirely manually created, during the prototyping of the NICE framework, including some particularly challenging cases pushing the limits of the framework's capabilities. Many were partially assisted by LLM tools, typical of modern software development practices (AI assistants such as Copilot or Codex). Finally, some recipes were generated fully automatically by LLM-based agents and later reviewed by humans before inclusion in the archive. This is the case in particular of the manually reviewed cases from the LLM pipeline presented in \Cref{sec:llm-evaluation}. Partially successful LLM cases can also contribute to expanding the archive after further refinement and human intervention.

\subsection{Versatility analysis} We review the archive's coverage in terms of weaknesses, attack vectors, software targets, and time.

\begin{figure}
    \centering
    \includegraphics[width=1.0\linewidth]{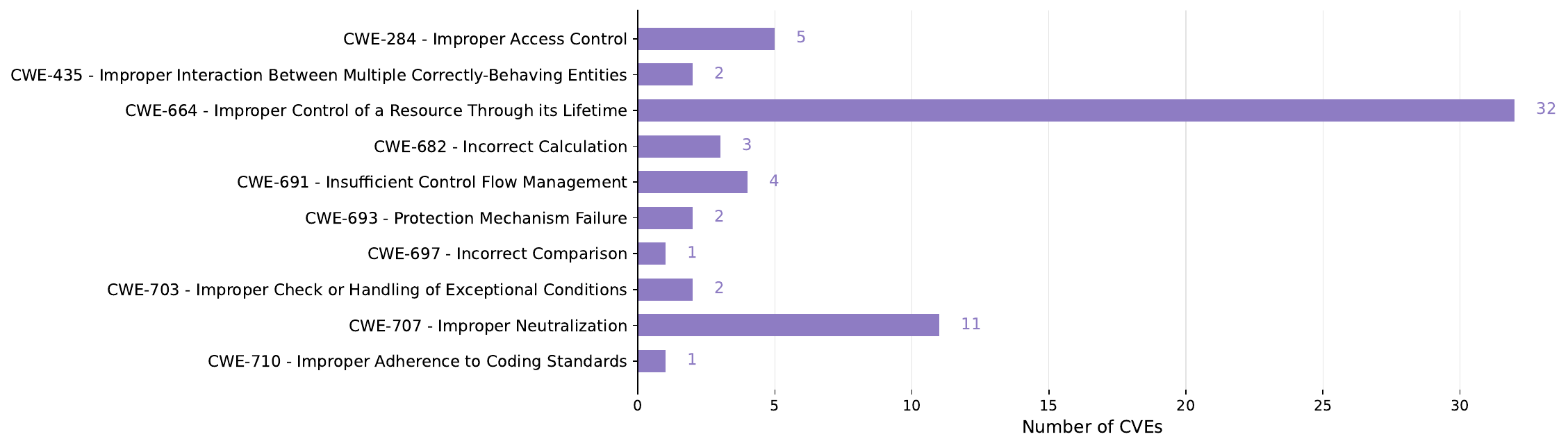}
    \caption{Distribution of CWE-1000 categories across reproduced CVEs. Note that the total exceeds the number of reproduced CVEs because some of them relate to multiple CWE categories.}
    \Description{Bar chart of the number of reproduced CVEs per top-level CWE-1000 pillar.}
    \label{fig:report-type-distri}
\end{figure}

\paragraph{Weakness.}
Figure~\ref{fig:report-type-distri} summarizes the weakness categories represented in our corpus using the CWE-1000 view~\cite{CWE1000View}, which maps specific CWE identifiers to ten top-level pillars. The spread across all pillars indicates that \textbf{NICE is broadly applicable to different vulnerability types}, rather than limited to a specific exploit primitive or family of weaknesses.

\paragraph{Attack vector.}
Our reproductions span network-facing attacks (e.g., Remote Code Execution), local attacks (e.g., privilege escalation), application-level attacks (e.g., Cross-Site Scripting, SQL Injection), and resource-exhaustion behaviors, indicating that \textbf{NICE accommodates heterogeneous exploitation scenarios}.

\begin{table}[t]
  \caption{Classification of vulnerability targets}
  \centering
  \footnotesize
  \setlength{\tabcolsep}{3pt}
  \renewcommand{\arraystretch}{1.15}
  \begin{tabular}{lL{0.55\columnwidth}c}
    \hline
    \textbf{Name} & \textbf{Description} & \textbf{Number of cases} \\
    \hline
    Headless application &
    Applications without a graphical interface, interacting via a command-line interface (CLI), library, or network protocol. & 49 \\
    System software &
    Software that depends on OS-level integration, e.g., system utilities. & 3 \\
    Kernel &
    Kernel, drivers, or modules that run in kernel space. & 3 \\
    Graphical application &
    Software with a GUI, requiring desktop sessions to run. & 1 \\
    \hline
  \end{tabular}
  \label{tab:vulnerability-categories}
\end{table}

\paragraph{Target software.}
Table~\ref{tab:vulnerability-categories} summarizes our taxonomy of vulnerability targets. We categorized the selected vulnerabilities into four software types: headless applications (49/\archiveSize), system software (3/\archiveSize), kernel (3/\archiveSize), and graphical applications (1/\archiveSize). This categorization reflects the degree of OS integration required for reproduction, and how different software types may require different ways to set up and install them in a NixOS environment.
Our reproductions cover all the categories, \textbf{showing that NICE achieves the objective of target software diversity}.

\begin{figure*}
    \centering
    \includegraphics[width=1.0\linewidth]{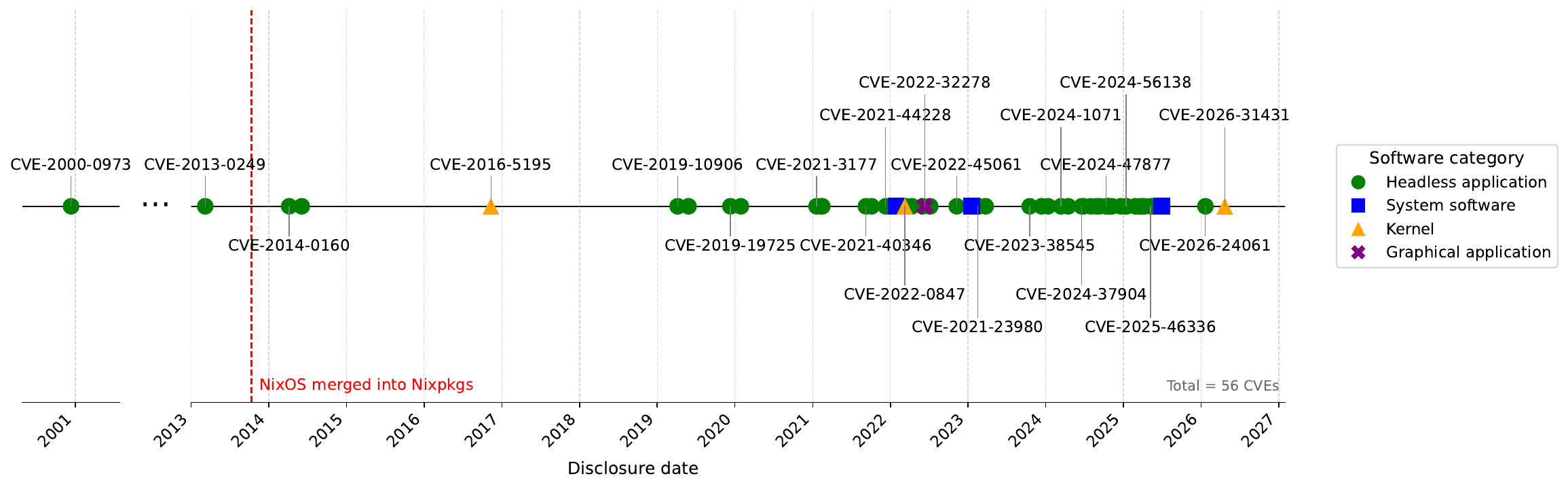}
    \caption{Timeline of CVEs reproduced using the NICE framework. The x-axis represents the CVE disclosure date.}
    \Description{Scatter plot of the reproduced CVEs over their disclosure dates, with marker shapes denoting the target software category.}
    \label{fig:report-timeline}
\end{figure*}

\paragraph{Temporal coverage.} We reproduced vulnerabilities ranging mostly from 2014 to 2026, but also including a few older cases (e.g., CVE-2000-0973, CVE-2013-0249), see Figure~\ref{fig:report-timeline}. This demonstrates that \textbf{NICE can reproduce and validate vulnerabilities over a long time span}.

\section{Automated Recipe Generation with LLM Agents}

\label{sec:llm-evaluation}

To demonstrate the feasibility of using NICE for vulnerability environment packaging, we conducted a lightweight evaluation of an agentic reproduction pipeline. The goal of this experiment is to assess whether the NICE documentation and tooling allow LLM agents to successfully create NICE recipes, and we do not aim to make a contribution in the field of LLM research itself.

\subsection{Dataset}

We use the ($D_L$) dataset released with CVE-GENIE~\cite{ullah2025cvegenie}, an automated vulnerability reproduction framework. The dataset contains 841 CVEs published between June 2024 and May 2025 that affect publicly available source-code projects, including CVEs without publicly available PoCs and CVEs targeting non-Linux platforms.
NICE is designed under the assumption that vulnerabilities to reproduce have proof-of-concept exploits publicly available, and is limited to the Linux platform.
Before the main experiment, we used 46 CVEs from ($D_L$) during a prototyping phase; these were excluded from the evaluation dataset. We then used an agent based on \texttt{nvidia/nemotron-3-ultra-550b-a55b} to filter the remaining CVEs and retain only Linux-targeted vulnerabilities with publicly available PoCs, resulting in 442 candidates. We randomly sampled from this pool until the predefined experimental budget was exhausted. In total, \LLMEvalSize\ CVEs were used for the LLM evaluation.

\subsection{Experimental setup}

The autonomous reproduction workflow takes a list of CVE identifiers as input. For each CVE, the orchestrator launches a NixOS 26.05 VM (8GB RAM, 80GB disk) and starts a reproducer agent.
The agent is implemented using the \texttt{OpenCode} agent harness, which provides the LLM with access to the working repository, shell commands, and the tools required to interact with the NICE environment. Each agent operates on a separate copy of the archive to avoid interference between concurrent reproductions.
Following the workflow from \Cref{sec:workflow}, the agent autonomously gathers information about the vulnerability and available public PoC, determines the required vulnerable software and configuration, implements the corresponding NICE recipe, iteratively performs exploitation and validation testing to refine the recipe, and documents the final result. At the end, a separate LLM agent evaluates the generated recipe by executing the validation test and reviewing the resulting artifacts before producing a final verdict. A subset of the successful recipes is then manually verified.

We impose several resource limits. Each CVE reproduction task has a global time limit of 90 minutes. Based on our prototyping phase, we additionally limit two potentially long-running activities to 40 minutes each: interactive experimentation with the VMs and implementation and refinement of the final validation test. These limits prevent the agent from spending most of its budget repeatedly modifying and rerunning commands, or waiting for commands that do not terminate. The remaining time is used for other activities such as vulnerability analysis, software packaging, NixOS configuration construction, and documentation. Each task is also limited to 400 LLM calls, 5 million tokens, and a cost of \$2. The orchestrator terminates the task when any limit is exceeded.

After prototyping with several LLMs, we selected \texttt{deepseek-\allowbreak v4-\allowbreak flash-\allowbreak 0731} for the experiment. It is a relatively affordable model while still providing sufficient agentic capability. All reproductions were started from the same version of the NICE framework and archive, including 23 examples.

\subsection{Results}

\begin{table*}[t]
  \centering
  \caption{Metrics for the LLM CVE reproduction experiment: average, minimum, and maximum values for LLM calls, input tokens, output tokens, cost in USD, and duration in minutes for each CVE reproduction task.}
  \label{tab:automated-results}
  \footnotesize
  \setlength{\tabcolsep}{6pt}
  \begin{tabular}{lccccc}
    \hline
    \textbf{Metric} &
    \textbf{LLM calls} &
    \textbf{Input tokens} &
    \textbf{Output tokens} &
    \textbf{Cost (\$)} &
    \textbf{Duration (min)} \\
    \hline
    Average & 227 & 692,703 & 47,416 & 0.73 & 62 \\
    Min & 6 & 83,812 & 1,257 & 0.04 & 13.5 \\
    Max (\emph{hard limit italicized}) & \emph{400} & 2,283,426 & 130,067 & 1.95 & \emph{90} \\
    \hline
  \end{tabular}
\end{table*}

Table~\ref{tab:automated-results} summarizes the metrics of the automated reproduction workflow. The average cost of each CVE reproduction task is \$0.73 and none of the tasks reached the \$2 limit. The average duration of each CVE reproduction task is 62 minutes and there are 34/\LLMEvalSize\ cases (30.9\%) that reached the 90-minute limit and were terminated by the orchestrator.

The reproducer claimed success in 58 of the 76 cases where it was not forcefully terminated.
The evaluator was then run on the outputs from all \LLMEvalSize\ cases.
It flagged no false positives, but did flag 2 cases (CVE-2024-36400 and CVE-2024-38514) where the reproducer produced a valid recipe, but was terminated by the orchestrator before reporting success.
The total of successful reproductions according to the evaluator is thus 60/\LLMEvalSize\ (\LLMEvalPercent).

We then manually reviewed 12 successful cases, including the two false negatives. Out of these, we found no case that was incorrectly marked as successful.
Successful evaluator verdicts always corresponded to meaningful reproductions that instantiated the vulnerable software, exercised the relevant behavior, and provided explicit validation evidence.

However, functional correctness does not imply that a recipe is optimal as a reusable artifact. We observed verbose documentation, overly detailed scenarios, and modeling choices that reduced realism or readability. For example, CVE-2024-38527, an XSS vulnerability in ZenUML, exposes attacker-related information directly to the server configuration. Similarly, CVE-2024-45596, a credential leak from a web cache, would benefit from a clearer separation between the victim, legitimate client, and attacker components.
Some recipes could use stronger or more representative assertions. We also observed a few tests that succeeded during evaluation and local reruns but failed in CI, suggesting environmental sensitivity or flakiness. Overall, the manual inspection provided a complementary view of recipe quality, highlighting areas for improvement in both the reproduction and evaluation process.

\subsection{Threats to validity}

\paragraph{Experimental non-determinism.}
The results may be affected by non-determinism in the LLM agents and by the experimental infrastructure. The same CVE could potentially lead to different recipes or outcomes across runs, while failures may also result from issues such as service-readiness races, CI behavior, or execution timeouts. In addition, the imposed limits on time, token usage, and LLM cost may terminate cases that could succeed with additional iterations.

\paragraph{Dataset construct validity.}
We used a first LLM agent to filter the dataset and exclude entries that did not target Linux-based systems or did not come with a public PoC.
This step may introduce biases or errors if the filtering agent incorrectly excludes relevant entries or includes irrelevant ones, potentially affecting the representativeness and completeness of the dataset.

\paragraph{Dataset external validity.}
The extracted dataset is limited to recent vulnerabilities and did not contain any kernel or system vulnerabilities, which limits the generalizability of our findings to these types of vulnerabilities.
Reproducing older vulnerabilities is typically more challenging and may not be achievable with a fully automated agentic pipeline at the low cost that we report.

\paragraph{Evaluator validity.}
The independent evaluator is itself an LLM and may incorrectly accept or reject generated recipes. In particular, it may misinterpret weak evidence, overlook implementation issues, or fail to execute an otherwise valid reproduction correctly. To reduce this risk, evaluator verdicts are based not only on textual inspection but also on execution of the generated tests, and a subset of results is manually reviewed. Nevertheless, the reported evaluator success rate should be understood as an automated approximation rather than perfect ground truth.

\paragraph{Generated artifact quality.}
LLM-generated recipes may contain unnecessarily verbose documentation, comments, or implementation details. While this generally does not affect the functional correctness of the reproduction, it can reduce readability and make the resulting artifacts harder to review and maintain. This affected most of the recipes we inspected. Such verbosity is partly a property of the generation model and prompting strategy rather than of NICE itself, and could be mitigated through stricter generation guidelines, post-processing, or human review.

Despite these limitations, the experiment provides evidence that NICE supports vulnerability reproduction across a range of real-world cases and gives an initial indication of the resources required for automated recipe construction.

\section{Discussion}
\label{sec:discussion}

We now reflect on what we learned from using the NICE framework to reproduce real-world vulnerabilities and, more generally, its design, limitations, and potential impact.

\subsection{Suitability of FPM for vulnerability reproduction}

Our experience suggests that FPM (in our case, Nix) is a good match for vulnerability reproduction, for three main reasons:
(1) \emph{Declarative system configuration} via NixOS allows expressing complex environments (services, networking, users, etc.) in a concise and readable way.
(2) \emph{The Nixpkgs package collection} covers a broad range of packages, making it likely that the vulnerable software and its dependencies are either already packaged or can be packaged with limited engineering effort.
(3) \emph{The reproducibility properties} of Nix ensure that environments can be rebuilt consistently over time and across different machines, which is crucial for maintaining vulnerability artifacts.

NICE builds on Nix's reproducibility properties, but its contribution is not the reproducibility mechanisms themselves. Rather, NICE defines how these mechanisms are used to package a vulnerability description as a complete executable blueprint. By leveraging pinned Nix inputs and historical package definitions in this way, NICE recipes can reproduce vulnerabilities years after their discovery—a significant advantage for long-term security research and training—while reducing several sources of environmental drift that commonly affect vulnerability artifacts over time.

\subsection{Exploit validation feasibility}

The manual process of verifying whether a vulnerability is present and exploitable can be fully automated with reasonable effort.
Once encoded as a NICE test, one can automatically validate if a given environment is indeed affected by a specific vulnerability, obtaining evidence of the expected security failure.
Automation then improves efficiency and reduces redundant, time-consuming manual testing work for all subsequent reproductions.
This matters in our setting because exploitation procedures are prone to error and oversight, especially in complex multi-host setups, and repeatedly performing the same steps by hand is tedious and makes results harder to compare across runs.

There are several key ingredients that make automated exploit validation significantly better than previous, more manual, reproduction approaches:
(1) \emph{Declarative test scripts} are not only programmable tests; they also document the exploitation procedure in a rigorous and auditable way.
(2) \emph{Provable vulnerability presence}, through tailored evidence and assertions that correspond to the exploit behavior, further strengthens the credibility of the reproduction.
(3) \emph{Reusable components} such as assertion blocks improve modularity and extensibility.
Similar to unit-testing frameworks, these components form a domain specific language that allows expressing concisely common evidence patterns that keep tests short and readable, while still making the security claim automatically verifiable.

These ingredients together create a self-documenting and convincing blueprint for vulnerability validation, enhancing auditability while providing explicit evidence of vulnerability presence and reducing maintenance effort. Additionally, tests can be run on non-vulnerable environments (e.g., patched versions) to provide negative controls and identify which configurations are necessary for successful exploitation.

The fact that vulnerability validation is restricted to a predefined set of assertions constrains how authors---and, in particular, AI agents---can establish that an exploit succeeded. They cannot simply introduce arbitrary exploit-specific checks whose success follows from behavior they themselves created; instead, the claimed security effect must be expressed through explicit, reusable validation primitives. This reduces the risk of tests being tailored to the expected outcome rather than to the vulnerability's actual security impact, which is something AI agents might be tempted to do because it is a shortcut to a successful test. This constrained and explicit validation model is a key focus of NICE. As LLM-assisted vulnerability discovery becomes more common~\cite{sheng2025llms}, it is increasingly important to ensure that resulting reproductions are genuine and supported by concrete, independently checkable evidence of the security failure.

NICE's automation capabilities fit naturally into continuous integration (CI) pipelines and work reliably across a wide range of vulnerability types and attack vectors.

\subsection{LLM-generated NICE recipes}
The LLM evaluation provides encouraging evidence that NICE recipes can be produced with limited manual effort. Given only a CVE identifier, the reproducer agent can perform much of the work: study vulnerability information and public PoCs, identify the required software and configuration, construct the environment, implement the exploitation scenario and validation test, and document the resulting recipe. Despite this broad task, the independent evaluator accepted 60 out of 110 generated recipes, suggesting that NICE, along with its accompanying documentation and tooling, provides a sufficiently structured target for translating vulnerability information into executable reproduction artifacts.

The resource requirements were also modest. We deliberately used the relatively inexpensive \texttt{deepseek-v4-flash-0731} model and limited each experiment to 90 minutes and \$2 of LLM cost. In practice, the average cost was only \$0.73 per CVE. These results show that useful recipes can already be generated without relying on an expensive frontier model or a large computational budget. Stronger models or larger resource budgets may further improve the level of automation, although evaluating this trade-off is outside the scope of this work.

Full autonomy is also not required for LLM assistance to be useful. Even when an agent does not complete a reproduction, it can still generate a substantial first draft of the NICE recipe, including vulnerable package selection, configuration, PoC adaptation, and test logic, which can then be reviewed and refined by a human. This can reduce the effort required to produce a high-quality reproduction.

Once a recipe has been completed and validated, the LLM is no longer required. The NICE artifact captures the vulnerable environment, exploitation procedure, and validation logic in an executable and reviewable form. The LLM cost is therefore mainly incurred during recipe creation, while subsequent reproductions remain inexpensive and independent of the generating model.

\subsection{Limitations}
While the framework worked across a diverse corpus, some limitations persist.

\paragraph{Exploit availability and robustness.}
Automated validation depends on PoCs that are reliable enough to run unattended. In practice, PoCs are not always publicly available, and when they are, they often lack information about prerequisites or make implicit assumptions, requiring trial and error to fill in the gaps.
However, the advantage of NICE is that once a vulnerable environment and its exploitation scenario have been successfully captured in a recipe, it continues to be reproducible and verifiable independently of the issues with the advisory description.

\paragraph{Scope boundaries.}
Our approach targets vulnerabilities that can be captured as software environments and exercised in virtualized systems. It does not directly address hardware vulnerabilities, and it is naturally biased toward software vulnerabilities exploitable in the Linux/NixOS context. OS-specific bugs on non-Linux platforms may require different execution platforms (e.g., a different operating system, or virtualization stack). Similarly, vulnerabilities that depend on physical interactions (e.g., side channels, hardware faults) are outside the scope of this approach.

\subsection{Potential impact}

\paragraph{Education and research.}
Cyber ranges are widely used in cybersecurity education to provide safe, isolated environments for hands-on practice, yet maintaining these labs over time is costly because images and setup instructions tend to drift and break. Our approach can reduce this burden by packaging each exercise as a declarative environment specification together with an automated validation test, so that cyber range operators can continuously check in CI/CD that a lab remains valid under pinned inputs, with the ability to roll back to the last known-good revision. Over time, a curated collection of such artifacts can serve as reusable teaching material and as a research corpus that preserves not only vulnerable software versions but also the concrete conditions required to reproduce and study them.

\paragraph{Vulnerability benchmarks.}
Automating the deployment of vulnerable environments is also key for the development and improvement of penetration testing tooling.
The same need (and advantages) of automatically deploying vulnerable environments for cyber ranges exists for tools.
If one wants to automatically test that some security tool is capable of detecting that a given set of target vulnerabilities is present in an environment, then they also need to be able to reliably deploy such an environment.
Short of that, automated testing of the tool will become flaky, which is undesirable.

The case of vulnerability benchmarks, when \emph{multiple} vulnerabilities need to be automatically tested, is a straightforward generalization of this scenario.

\paragraph{Improving vulnerability reporting and triage.}
The declarative nature of NICE makes it a good fit for enhancing vulnerability reporting.
If a vulnerability report is submitted accompanied by a NICE recipe, its reviewers and triagers can examine a concise and executable recipe rather than relying solely on textual descriptions.
The ability of test scripts to provide provable evidence of vulnerability presence further strengthens the credibility of the reproduction.
This becomes particularly relevant now that many vulnerabilities are reported by or with the help of AI tools.

With this framework, bounty programs and security researchers can set up automated pipelines to validate reported vulnerabilities consistently.
A reporter would not only claim that a vulnerability exists with observable effects, but would also need to provide specific tests demonstrating that an illegal state was reached after performing the exploit.

\subsection{Future directions}
\label{sec:future-directions}
Several directions would strengthen the approach and broaden its impact.

 \paragraph{Expanding beyond NixOS} While NixOS provides a solid foundation, extending support to other Linux distributions or even non-Linux systems would increase the framework's applicability. Particularly, for OS-specific vulnerabilities, widening the scope or allowing the validation of exploits on different OSes would be beneficial. This could be achieved while still relying on the NixOS test framework and integrating it with VM images for other OSes.
\paragraph{A collaborative archive of vulnerability recipes.} NICE already provides a public archive of vulnerability reproduction recipes and tests. A natural next step is to make this archive collaborative, allowing enthusiasts to contribute new recipes, improve or fix existing ones over time. Such a community-maintained repository could enlarge the corpus of reproducible vulnerabilities, and provide a valuable database to the security community.
\paragraph{Scaling up the corpus.} Building a larger collection of vulnerability reproductions would enable more systematic measurement of what makes vulnerabilities reproducible, what breaks over time, and which classes of issues benefit most from FPM. A sizable dataset of declarative vulnerability recipes would also assist in training and evaluating AI models for security tasks, as this addresses the current lack of high-quality, machine-readable advisories.

\section{Related Work}
\label{sec:related}

In this section, we compare our work with the related literature. First, we compare to other tools for reproducing vulnerabilities. Next, we compare to other uses of Functional Package Managers for reproducibility use cases.

\subsection{Reproducing software vulnerabilities}

We review tools focused on reproducing software vulnerabilities. As their design objectives vary greatly, we group them by goal.

\paragraph{Reproducing vulnerable environments}
Akasaka et al.~\cite{akasaka2020reproducible} and Kamata et al.~\cite{kamata2024Intention} pioneered the use of Infrastructure as Code (IaC) to formalize vulnerable environments and validate them automatically. Their tools, \textit{vultest} and the follow-up \textit{vulen},\footnote{\textit{vulen} is not publicly available, and despite our attempts to contact the authors, we were unable to access it.} use structured YAML files to provide human-readable environment descriptions and automated exploit validation. However, their reliance on default package managers and manually maintained base images reduces long-term reproducibility as repositories and images change. In our evaluation of \textit{vultest}, most vulnerabilities could not be reproduced because the base images were inaccessible, and substituting alternatives further undermined reproducibility. Moreover, validation is limited to a small set of tools (Metasploit~\cite{metasploit} and HTTP requests), restricting custom exploit techniques and thus not fully satisfying our validation objective.

Similarly, the Debian CVE Reproduction Tool (DECRET)~\cite{decret} automates container creation using Debian snapshot services to obtain historical package versions. This approach ensures the availability of vulnerable software. Crucially, its reliance on containers prevents the reproduction of kernel vulnerabilities.

Vulhub~\cite{vulhub}, a community effort, provides a large collection of ready-to-run vulnerable Docker scenarios for security education and experimentation. Its long-term stability however depends on continued maintenance of Dockerfiles, upstream base images, and external download URLs. Additionally, users cannot easily inspect or modify the underlying environment, which limits applicability and extensibility.

KernJC~\cite{ruan2024kernjc} specifically targets kernel vulnerabilities by providing a framework to reproduce and provision them using QEMU virtual machines.

Caturano et al.~\cite{exploitwp2docker} propose Exploit2Docker, a platform that automatically generates WordPress vulnerability containers. It targets solely WordPress vulnerabilities, and successfully generates 484 scenarios of WordPress vulnerabilities. However, this does not mean that these containers are exploitable, as no validation test is provided.

NICE addresses the limitations raised by these tools. (1) Vulhub, \textit{vultest}, and Exploit2Docker rely on mutable base images and upstream registries, meaning their artifacts may degrade as those dependencies evolve. NICE inherits key properties from Nix, including reproducible recipes, component pinning, and access to the public binary cache. Empirical research shows that 99.5\% of Nix packages can be successfully rebuilt more than five years later~\cite{malka:hal-04913007}, supporting NICE's long-term reproducibility objective. (2) NICE supports a wide range of software types, including kernel vulnerabilities, which are not reproducible with container-based approaches such as DECRET, Vulhub, and Exploit2Docker. (3) NICE provides automated validation tests to ensure that reproduced environments are indeed vulnerable and exploitable, addressing a limitation of approaches such as Exploit2Docker.

\paragraph{Automating the reproduction of vulnerabilities}
Recent advancements in large language models (LLMs) have led to the development of agent-based approaches in various domains of computer science, including security tasks such as exploit generation~\cite{shen2025pentestagentincorporatingllmagents,liu2024refiningchatgpt}.
Recently, LLMs have also been explored for vulnerability reproduction. For instance, Ullah et al.~\cite{ullah2025cvegenie} introduced CVE-GENIE, an automated multi-agent framework that can reproduce real-world vulnerabilities given their CVE identifiers. CVE-GENIE demonstrates the potential of LLMs in vulnerability reproduction and validation with a very large set of CVEs and fully automated processes, achieving a reproduction rate of 51\% on 841 CVEs at an estimated cost of \$2.77 per CVE.

The idea of using LLMs to generate vulnerable environments is complementary to our approach, as we have shown in \Cref{sec:llm-evaluation}. LLMs can be used to assist in generating vulnerability recipes that are then stabilized and validated within our framework.
While approaches such as CVE-GENIE rely on LLMs every time a vulnerable environment needs to be reproduced, NICE only (optionally) requires the LLM during the initial recipe generation. Once the recipe is created, it can be reused indefinitely without further LLM involvement.
Furthermore, the proof of exploitation is a declarative, textual artifact that is reviewable by a human. It does not require trusting the LLM's claims.

Other works also try to leverage LLMs with similar high-level structures for automating vulnerability reproduction and validation, but evaluate on smaller case studies~\cite{lotfi2025automatedvulnerabilityvalidationverification} or on cybersecurity benchmarks for evaluating agent capabilities~\cite{zhu2025cvebenchbenchmarkaiagents, zhang2025cybenchframeworkevaluatingcybersecurity}.

\paragraph{Creating vulnerable environments}
Other projects also aim to reproduce vulnerabilities for more specific use cases.
For the use case of building cyber ranges, CyExec*~\cite{cyexecstar}, CyRIS~\cite{cyris}, and SecGen~\cite{secgen} focus on automating the generation of vulnerable environments and facilitating their deployment for training and simulation purposes. These tools resemble our framework in creating vulnerable environments, but they differ in design objectives: they focus on synthesized vulnerabilities for education purposes, rather than faithfully reproducing and validating real-world vulnerabilities.

\subsection{FPM for reproducibility}

On the front of reproducibility in FPM, Malka et al.~performed the first large-scale empirical study of bitwise reproducibility, showing that 91\% of the packages in the \textit{nixpkgs} distribution are bitwise reproducible~\cite{malka:hal-04913007}. In another work~\cite{NixSpaceTime} they also showed that \textit{Nix} can reproduce past build environments of about 7 million packages from 200 historical revisions of \textit{Nixpkgs} reliably.

Hausch et al.~\cite{hausch2025preciceeco} demonstrate the benefits of Nix and NixOS for scientific computing by packaging the \emph{preCICE} ecosystem---an open-source coupling framework for multi-physics simulations. Despite some non-packageable components, they conclude that Nix is suitable for reproducible software environments. Inspired by these results, we investigated applying the same principles to vulnerability reproduction, where preserving exact software stacks is essential.

\section{Conclusion}
\label{sec:conclusion}

In this paper we introduced the NICE framework to reproduce vulnerable environments and automatically validate the presence of exploitable vulnerabilities within them.
The framework defines a systematic approach to construct NixOS virtual machines that contain vulnerable software and its dependencies.
NixOS's declarative configuration and reproducibility properties make it a good fit for vulnerability reproduction, as well as audits by third parties.
NICE also addresses the challenge of validating whether a vulnerability is indeed present by automating the validation process and providing factual evidence of exploitability (or the equivalent of a ``test failure'' otherwise).

We contribute a public archive of \archiveSize\ reproducible vulnerability recipes, covering different types of software, ranging from userland and graphical applications to system software and kernel vulnerabilities. The reproduced vulnerabilities span many CWE categories and attack vectors and cover an extensive time period. We also evaluate using LLM agents to automatically generate NICE recipes. We achieve a fully automated reproduction rate of \LLMEvalPercent\ on a set of \LLMEvalSize\ CVEs, demonstrating that the framework is suitable for AI-assisted vulnerable environment packaging.

\section*{Data Availability}
A complete replication package for this work is available, the replication package is in two parts:
\begin{enumerate}

\item The current version of the NICE archive, including all reproduced vulnerability recipes, tools, and documentation for the NICE framework~\cite{nice-archive-github}.

\item A snapshot of the NICE archive used in the experiment, the LLM evaluation dataset, generated NICE recipes, independent evaluator reports, and metadata and tools for generating the charts~\cite{this-replication-package-zenodo}.

\end{enumerate}

\bibliographystyle{ACM-Reference-Format}
\bibliography{nicearchive}

\end{document}